\documentclass[12pt]{article}
\evensidemargin\oddsidemargin
\def\text#1{\mbox{ #1}\,}
\newtheorem{definition}{Definition}

\begin{document}

\bigskip

\begin{center}
{\LARGE Exactly soluble models of decoherence\footnote{Extended
version of a talk presented at the 7th UK Conference on Mathematical and
Conceptual Foundations of Modern Physics, Nottingham 7 - 11 September 1998}}

\bigskip

{\large Joachim Kupsch\footnote{e-mail: kupsch@physik.uni-kl.de}}

{\large Fachbereich Physik, Universit\"at Kaiserslautern\\D-67653
Kaiserslautern, Germany}

\medskip
\end{center}

\begin{abstract}
Superselection rules induced by the interaction with the environment are
investigated with the help of exactly soluble Hamiltonian models. Starting
from the examples of Araki and of Zurek more general models with scattering
are presented for which the projection operators onto the induced
superselection sectors do no longer commute with the Hamiltonian. The
example of an environment given by a free quantum field indicates that
infrared divergence plays an essential role for the emergence of induced
superselection sectors. For all models the induced superselection sectors
are uniquely determined by the Hamiltonian, whereas the time scale of the
decoherence depends crucially on the initial state of the total system.
\end{abstract}

\section{Introduction}

One of the puzzles of quantum mechanics is the question, how classical
objects can arise in quantum theory. Quantum mechanics is a statistical
theory, but its statistics differs on a fundamental level from the
statistics of classical objects. The violation of Bell's inequalities and
the context dependence of quantum mechanics (Kochen-Specker theorem)
illustrate this fact, see e.g. \cite{Peres:1995}.

It is known since a long time that the statistical results of quantum
mechanics become consistent with a classical statistics of ``facts'', if the
superposition principle is reduced to ``superselection sectors'', i.e.
coherent orthogonal subspaces of the full Hilbert space. The mathematical
structure of quantum mechanics and of quantum field theory provides us with
only a few ``superselection rules'', the most important being the charge
superselection rule related to gauge invariance, see e.g. \cite{BLOT:1990} 
\cite{Wightman:1995} and the references given therein. But there are
definitively not enough of these superselection rules to understand
classical properties in quantum theory. A possible solution of this problem
is the emergence of effective superselection rules due to decoherence caused
by the interaction with the environment. These investigations -- often
related to a discussion of the process of measurement -- have developed in
the eighties; some references are \cite{Araki:1980}\cite{Zurek:1982}\cite
{Joos/Zeh:1985}, but see also the earlier publications \cite{Zeh:1970}\cite
{Zeh:1971} and \cite{Emch:1972a}.

In this article decoherence and the emergence of environment induced
superselection rules are investigated on the basis of exactly soluble
models. After a short introduction to superselection rules and to the
dynamics of subsystems in Sects. \ref{ssr} and \ref{subsys}, several models
are presented in Sect. \ref{models}. For a class of simple models, which
essentially go back to Araki \cite{Araki:1980} and Zurek \cite{Zurek:1982},
the transition between the induced superselection sectors is suppressed
uniformly in trace norm. In a more realistic example with a quantum field as
environment, presented in Sect. \ref{free}, the infrared behaviour of the
environment is of essential importance for the emergence of induced
superselection rules. Here uniform estimates, which persist for arbitrary
times, are only possible in the limit of infrared divergence. In Sect. \ref
{scatt} it is shown that additional scattering processes (by sufficiently
smooth potentials) do not alter the induced superselection sectors, but the
decoherence is no longer uniform with respect to the initial state of the
system.

\section{Superselection rules\label{ssr}}

We start with a few mathematical notations. Let $\mathcal{H}$ be a separable
Hilbert space, then the following spaces of linear operators are used.

$\mathcal{B}(\mathcal{H})$: The $\mathbf{R}$-linear space of all bounded
self-adjoint operators $A$. The norm of this space is the operator norm $%
\Vert A\Vert $.

$\mathcal{T}(\mathcal{H})$: The $\mathbf{R}$-linear space of all
self-adjoint nuclear operators $A$. These operators have a pure point
spectrum ${\alpha _i}\in \mathbf{R},\,i=1,2,...,$ with $\sum_i|\alpha
_i|<\infty $. The natural norm of this space is the trace norm $\Vert A\Vert
_1=\mathrm{tr}\sqrt{A^{+}A}=\sum_i|\alpha _i|$. Another norm, used in the
following sections, is the Hilbert-Schmidt norm $\Vert A\Vert _2=\sqrt{%
\mathrm{tr}\,A^{+}A}$. These norms satisfy the inequalities $\Vert A\Vert
\leq \Vert A\Vert _2\leq \Vert A\Vert _1$.

$\mathcal{D}(\mathcal{H})$: The set of all statistical operators, i.e.
positive nuclear operators $W$ with a normalized trace, $\mathrm{tr}\,W=1 $.

$\mathcal{P}(\mathcal{H})$: The set of all rank one projection operators $%
P^1 $.

\noindent These sets satisfy the obvious inclusions $\mathcal{P}(\mathcal{H}%
)\subset \mathcal{D}(\mathcal{H})\subset \mathcal{T}(\mathcal{H})\subset 
\mathcal{B}(\mathcal{H}).$

Any state of a quantum system is represented by a statistical operator $W\in 
\mathcal{D}(\mathcal{H})$, the elements of $\mathcal{P}(\mathcal{H})$
thereby correspond to the pure states. Any (bounded) observable is
represented by an operator $A\in \mathcal{B}(\mathcal{H})$, and the
expectation of the observable $A$ in the state $W$ is the trace $\mathrm{tr}%
\,WA$. Without additional knowledge about the structure of the system we
have to assume that the set of all states corresponds exactly to $\mathcal{D}%
(\mathcal{H})$, and the set of all (bounded) observables is $\mathcal{B}(%
\mathcal{H})$. The state space $\mathcal{D}(\mathcal{H})$ has an essential
property: it is a convex set, i.e. $W_1,W_2\in \mathcal{D}(\mathcal{H})$
implies $\lambda _1W_1+\lambda _2W_2\in \mathcal{D}(\mathcal{H})$ if $%
\lambda _{1,2}\geq 0$ and $\lambda _1+\lambda _2=1.$ Any statistical
operator $W\in \mathcal{D}(\mathcal{H})$ can be decomposed into pure states $%
W=\sum_nw_nP_n^1$ with $P_n^1\in \mathcal{P}(\mathcal{H})$ and probabilities 
$w_n\geq 0,\;\sum_nw_n=1$. An explicit example is the spectral decomposition
of $W$. But there are many other possibilities. It is exactly this
arbitrariness that does not allow a classical interpretation of quantum
probability. A more detailed discussion of the state space of quantum
mechanics can be found in \cite{Kupsch:1996a}.

The arbitrariness of the decomposition of $W$ originates in the
superposition principle. In quantum mechanics, especially in quantum field
theory, the superposition principle can be restricted by superselection
rules. Here we cannot discuss the arguments to establish such rules, for
that purpose see e.g. \cite{BLOT:1990}\cite{Wightman:1995} and also Chap.6
of \cite{GJKKSZ:1996}, or to refute them, see e.g. \cite{Mirman:1979}. Here
we only investigate the consequences for the structure of the state space.
In a theory with discrete superselection rules like the charge
superselection rule, the Hilbert space $\mathcal{H}$ splits into orthogonal
superselection sectors $\mathcal{H}_m,\,m\in \mathbf{M,}$ such that $%
\mathcal{H=}\oplus _m\mathcal{H}_m$. Pure states with charge $m$ (in
appropriate normalization) are then represented by vectors in $\mathcal{H}_m$%
, and superpositions of vectors with different charges have no physical
interpretation. The projection operators $P_m$ onto the orthogonal subspaces 
$\mathcal{H}_m$ satisfy $P_mP_n=\delta _{mn}$ and $\sum_mP_m=I$. The set of
states is reduced to those statistical operators which satisfy $P_mW=WP_m$
for all projection operators $P_m,m\in \mathbf{M}$. The state space of the
system is then $\mathcal{D}^S=\{W\in \mathcal{D}(\mathcal{H}%
)|WP_m=P_mW,\,m\in \mathbf{M}\}$, and all statistical operators satisfy the
identity $W=\sum_mP_mWP_m$. An equivalent statement is that all observables
of such a system have to commute with the projection operators $P_m,\,m\in 
\mathbf{M,}$ and the set of observables of the system is given by \\$%
\mathcal{B}^S=\left\{ A\in \mathcal{B}(\mathcal{H})\mid AP_m=P_mA,\,m\in 
\mathbf{M}\right\} =\left\{ A\in \mathcal{B}(\mathcal{H})\mid
A=\sum_mP_mAP_m\right\} .$

The projection operators $\{P_m\mid m\in \mathbf{M}\}$ are themselves
observables, which commute with all observables of the system, and they
generate a nontrivial centre of the algebra of observables.

In theories with continuous superselection rules the finite or countable set
of projection operators $\left\{ P_m,m\in \mathbf{M}\right\} $ is
substituted by a (weakly continuous) family of projection operators $%
P(\Delta )$ indexed by measurable subsets $\Delta \subset \mathbf{R}$, see
e.g. \cite{Piron:1969} or \cite{Araki:1980}. These projection operators have
to satisfy 
\begin{equation}
\left\{ 
\begin{array}{l}
P(\Delta _1\cup \Delta _2)=P(\Delta _1)+P(\Delta _2)%
\mbox{  for all
intervalls }\,\Delta _1,\Delta _2 \\ 
P(\Delta _1)P(\Delta _2)=O\mbox{ \ if\ }\,\Delta _1\cap \Delta _2=\emptyset ,%
\mbox{  and }\,P(\emptyset )=O,\;P(\mathbf{R})=1.
\end{array}
\right.  \label{sp}
\end{equation}
$\;$The set of observables is now given by $\mathcal{B}^S=\left\{ A\in 
\mathcal{B}(\mathcal{H})\mid AP(\Delta )=P(\Delta )A,\,\Delta \subset 
\mathbf{R}\right\} $, but there is no formulation of the corresponding set
of states within the class of nuclear statistical operators.

The importance of superselection rules for the transition from quantum
probability to classical probability is obvious. But there remains an
essential problem: Only very few superselection rules can be found in
quantum mechanics that are compatible with the mathematical structure and
with experiment. A satisfactory solution to this problem is the emergence of
effective superselection rules induced by the interaction with the
environment.

\section{Dynamics of subsystems and induced superselection sectors\label
{subsys}}

In the following we consider an ``open system'', i.e. a system $S$ which
interacts with an ``environment'' $E$, such that the total system $S+E$
satisfies the usual Hamiltonian dynamics. The Hilbert space $\mathcal{H}%
_{S+E}$ of the total system $S+E$ is the tensor space $\mathcal{H}_S\otimes 
\mathcal{H}_E$ of the Hilbert spaces for $S$ and for $E$. We assume that the
only observables at our disposal are the operators $A\otimes I_E$ with $A\in 
\mathcal{B}(\mathcal{H}_S)$. If the state of the total system is $W\in 
\mathcal{D}(\mathcal{H}_{S+E})$, then all expectation values $\mathrm{tr}%
_{S+E}W(A\otimes I_E)$ can be calculated from the reduced statistical
operator $\rho =\mathrm{tr}_EW$ which is an element of $\mathcal{D}(\mathcal{%
H}_S)$, such that $\mathrm{tr}_SA\rho =\mathrm{tr}_{S+E}(A\otimes I_E)W$
holds for all $A\in \mathcal{B}(\mathcal{H}_S)$. We shall refer to the
statistical operator $\rho =\mathrm{tr}_EW$ as the ``state'' of the
subsystem.

As mentioned above we assume the usual Hamiltonian dynamics for the total
system, i.e. $W(t)=U(t)WU^{+}(t)$ with the unitary group $U(t)$, generated
by the total Hamiltonian. Except for the trivial case that $S$ and $E$ do
not interact, the dynamics of the reduced statistical operator 
\begin{equation}
\rho (t)=\mathrm{tr}_EU(t)WU^{+}(t)  \label{ss.7}
\end{equation}
is no longer unitary, and it is exactly this dynamics which can produce
effective superselection sectors. More explicitly, the Hamiltonian of the
total system can provide a family of projection operators $\left\{
P_m,\,m\in \mathbf{M}\right\} $ which are independent from the initial
state, such that the statistical operator behaves like 
\begin{equation}
\rho (t)\cong \sum_mP_m\rho (t)P_m\;\mbox{  for }\,t\rightarrow \infty .
\label{ss.8}
\end{equation}
An equivalent statement is that the superpositions between vectors of
different sectors $P_m\mathcal{H}_S$ are strongly suppressed. Any mechanism,
which leads to this effect, will be called \textit{decoherence}.

In the case of induced continuous superselection rule the asymptotics is
more appropriately described in the Heisenberg picture, as stated above. But
the decoherence effect is also seen in the Schr\"odinger picture: $P(\Delta
_1)\rho (t)P(\Delta _2)\rightarrow 0$ for $t\rightarrow \infty $ if $\Delta
_1$ and $\Delta _2$ have a positive distance.

The statement (\ref{ss.8}) is so far rather vague since it does not specify
the asymptotics. A preliminary definition of a \textit{weak} type of
decoherence can be formulated as follows.

\begin{definition}
\label{ss} The subspaces $P_m\mathcal{H}_S,\,m\in \mathbf{M,}$ are denoted
as induced superselection sectors, of the dynamics (\ref{ss.7}), if for all
observables $A\in \mathcal{B}(\mathcal{H}_S)$ which have no diagonal matrix
elements, i.e. $P_mAP_m=O,\,m\in \mathbf{M,}$ the trace 
\begin{equation}
\mathrm{tr}_{S+E}(A\otimes I_E)U(t)WU^{+}(t)=\mathrm{tr}_SA\rho (t)
\label{ss.9}
\end{equation}
vanishes if $t\rightarrow \infty $ for all initial states $W\in \mathcal{D}_1
$ of a dense subset $\mathcal{D}_1\subset \mathcal{D}(\mathcal{H}_{S+E})$.
\end{definition}

It is possible to give an alternative definition with $\mathcal{D}_1$
substituted by $\mathcal{D}(\mathcal{H}_{S+E})$. These definitions are
equivalent, as can be easily seen. Assume the statements of Definition \ref
{ss} are valid for a family of subspaces $\left\{ P_m\mathcal{H}_S,\,m\in 
\mathbf{M}\right\} $, then we can find for any $W\in \mathcal{D}(\mathcal{H}%
_{S+E})$ and any $\varepsilon >0$ a statistical operator $W_1\in \mathcal{D}%
_1$ such that $\left\| W-W_1\right\| _1<\varepsilon $ and \\ $\mathrm{tr}%
_{S+E}(A\otimes I_E)U(t)W_1U^{+}(t)\rightarrow 0$ if $t\rightarrow \infty $
for the specified class of observables $A$. Since \\ $\left| \mathrm{tr}%
_{S+E}(A\otimes I_E)U(t)(W-W_1)U^{+}(t)\right| <\varepsilon \left\|
A\right\| $ the trace (\ref{ss.9}) vanishes if $t\rightarrow \infty $ for
all initial states $W\in \mathcal{D}(\mathcal{H}_{S+E})$.

The independence from the initial state justifies the terminology induced
''superselection'' rules. The Definition \ref{ss} has to be supplemented by
statements about the time scale of the convergence. For that purpose the
following models are investigated. They indicate the essential role of the
initial state -- especially of the components affiliated to the environment
-- to achieve decoherence in sufficiently short time.

\section{Soluble models\label{models}}

The first class of the presented models has a discrete superselection
structure such that the off-diagonal elements of the statistical operator
vanish in trace norm $\left\| .\right\| _1$%
\begin{equation}
\left\| P_m\rho (t)P_n\right\| _1\rightarrow 0\mbox{  if }\,t\rightarrow
\infty \mbox{  and }\,m\neq n  \label{mod.0}
\end{equation}
for an arbitrary initial state $\rho (0)\in \mathcal{D}(\mathcal{H}_S)$. But
the asymptotics is more complicated for the more realistic models
investigated in Sects. \ref{free} and \ref{scatt}.

The models of Sects. \ref{Araki} and \ref{free} have the following
structure. The Hilbert space is $\mathcal{H}_{S+E}=\mathcal{H}_S\otimes 
\mathcal{H}_E.$ The total Hamiltonian has the form 
\begin{equation}
H_{S+E}=H_S\otimes I_E+I_S\otimes H_E+V_S\otimes V_E  \label{mod.2}
\end{equation}
where $H_S$ is the Hamiltonian of S, $H_E$ is the Hamiltonian of E, $%
V_S\otimes V_E$ is the interaction term between S and E with self-adjoint
operators $V_S$ on $\mathcal{H}_S$ and $V_E$ on $\mathcal{H}_E$. We make the
following assumptions

\begin{enumerate}
\item[1)]  The operators $H_S$ and $V_S$ commute, $\left[ H_S,V_S\right] =O,$
hence $\left[ H_S\otimes I_E,V_S\otimes V_E\right] =O.$

\item[2)]  The operator $V_E$ has an absolutely continuous spectrum.
\end{enumerate}

The assumption 1) is a rather severe restriction, which will be given up in
Sect. \ref{scatt}, where we admit an additional scattering potential $V$,
which has not to commute with any of the other operators. The assumption 2)
has more technical reasons. It implies that estimates can be derived in the
limit $t\rightarrow \infty $ in agreement with Definition \ref{ss}. But one
can also allow operators with point spectra (as done in \cite{Zurek:1982}),
if the spacing of the eigenvalues is sufficiently small. Then the norm in (%
\ref{mod.0}) is an almost periodic function, and the suppression of this
norm takes place only during a finite time interval $0\leq t\leq T$. But $T$
can be large enough for all practical purposes.

The operator $V_S$ has the spectral representation $V_S=\int_{\mathbf{R}%
}\lambda P(d\lambda )$ with a spectral family $\left\{ P(\Delta ),\,\Delta
\subset \mathbf{R}\right\} $ which satisfies (\ref{sp}). We shall see that
exactly this spectral family determines the superselection sectors. If $V_S$
has a pure point spectrum, then $P(\Delta )$ is a step function with values $%
P_m$, and we can write 
\begin{equation}
V_S=\sum_m\lambda _mP_m.  \label{mod.5}
\end{equation}
As a consequence of assumption 1) we have $\left[ H_S,P(\Delta )\right] =O$
or $\,\left[ H_S,P_m\right] =O$ for $\Delta \subset \mathbf{R}$ or $m\in 
\mathbf{M}$, respectively. The Hamiltonian (\ref{mod.2}) has therefore the
form (for simplicity we only write the version with the discrete spectrum (%
\ref{mod.5})) 
\begin{eqnarray}
H_{S+E} &=&H_S\otimes I_E+\sum_mP_m\otimes \Gamma _m\mbox{  with}\,
\label{mod.7} \\
\Gamma _m &=&H_E+\lambda _mV_E.  \label{mod.8}
\end{eqnarray}
The unitary evolution $U(t):=\exp (-iH_{S+E}t)$ of the total system can be
written as \\$\left( \mathrm{e}^{-\imath H_St}\otimes I_E\right)
\sum_mP_m\otimes \mathrm{e}^{-i\Gamma \lambda _mt}$. The calculation of the
reduced dynamics (\ref{ss.7}) then leads to 
\begin{equation}
P_m\rho (t)P_n=P_m\mathrm{e}^{-\imath H_St}\left( \mathrm{tr}_E\mathrm{e}%
^{-i\Gamma _mt}W\mathrm{e}^{i\Gamma _nt}\right) \mathrm{e}^{\imath H_St}P_n,
\label{mod.10}
\end{equation}
where the operators $P_n$ are the projection operators of the spectral
representation (\ref{mod.5}) of $V_S$. For a factorizing initial state $%
W=\rho \otimes \omega $ with $\rho \in \mathcal{D}(\mathcal{H}_S)$ and a
reference state $\omega \in \mathcal{D}(\mathcal{H}_E)$ of the environment,
the operator (\ref{mod.10}) simplifies to $P_m\rho (t)P_n=P_me^{-iH_St}\rho
e^{iH_St}P_n\,\chi _{m,n}(t)$ with 
\begin{equation}
\chi _{m,n}(t)=\mathrm{tr}_E\left( \mathrm{e}^{i\Gamma _nt}\mathrm{e}%
^{-i\Gamma _mt}\omega \right)  \label{mod.13}
\end{equation}
and the emergence of dynamically induced superselection rules depends on an
estimate of this trace.

\subsection{The Araki-Zurek models\label{Araki}}

The first soluble models for the investigation of the reduced dynamics have
been given by Araki \cite{Araki:1980} and Zurek \cite{Zurek:1982}, and the
following construction is essentially based on these papers. In addition to
the specifications made above, we demand that

\begin{enumerate}
\item[3)]  the Hamiltonian $H_E$ and the potential $V_E$ commute, $\left[
H_E,V_E\right] =O.$
\end{enumerate}

We first investigate $P_m\rho (t)P_n$ for a factorizing initial state $%
W=\rho \otimes \omega $. Under the assumption 3) the trace (\ref{mod.13})
simplifies to $\chi _{m,n}(t)=\mathrm{tr}_E\left( \mathrm{e}^{-i(\lambda
_m-\lambda _n)V_Et}\omega \right) $. Let $V_E=\int_{\mathbf{R}}\lambda
P_E(d\lambda )$ be the spectral representation of the operator $V_E$. Then,
as a consequence of assumption 2), for any $\omega \in \mathcal{D}(\mathcal{H%
}_E)$ the measure $d\mu (\lambda ):=\mathrm{tr}_E\left( P_E(d\lambda
)\,\omega \right) $ is absolutely continuous with respect to the Lebesgue
measure, and the function $\chi (t):=\mathrm{tr}\left( \mathrm{e}^{-\imath
V_Et}\omega \right) =\int_{\mathbf{R}}\mathrm{e}^{-i\lambda t}$ $d\mu
(\lambda )$ vanishes if $t\rightarrow \infty $. But to have a decrease which
is effective in sufficiently short time, we need an additional smoothness
condition on $\omega $ (which does not impose restrictions on the
statistical operator $\rho \in \mathcal{D}(\mathcal{H}_S)$ of the system S).
If the integral operator, which represents $\omega $ in the spectral
representation of $V_E$, is a sufficiently differentiable function
(vanishing at the boundary points of the spectrum) we can derive estimates
like $\left| \chi (t)\right| \leq C_\gamma (1+\left| t\right| )^{-\gamma }$
with arbitrarily large values of $\gamma $. Such an estimate leads to the
upper bound 
\begin{equation}
|\chi _{m,n}(t)|\leq C_\gamma (1+\delta \left| t\right| )^{-\gamma }
\label{mod.17}
\end{equation}
if $\left| \lambda _m-\lambda _n\right| \geq \delta >0$, and we obtain an
estimate for the norm (\ref{mod.0}) 
\begin{equation}
\left\| P_m\rho (t)P_n\right\| _1\leq C_\gamma (1+\delta \left| t\right|
)^{-\gamma }.  \label{mod.1}
\end{equation}
with arbitrary $\rho (0)\equiv \rho \in \mathcal{D}(\mathcal{H}_S)$. The
constants $\gamma >0,$ $\delta >0$ and $C_\gamma >0$ do not depend on $\rho $%
. Moreover one can achieve large values of $\gamma $ and/or small values of
the constant $C_\gamma $ if the reference state $\omega $ is sufficiently
smooth.

These results depend on the reference state $\omega $ only via the decrease
of $\chi (t)$. We could have chosen a more general initial state $W\in 
\mathcal{D}(\mathcal{H}_{S+E})$%
\begin{equation}
W=\sum_\mu c_\mu \,\rho _\mu \otimes \omega _\mu  \label{mod.18}
\end{equation}
with $\rho _\mu \in \mathcal{D}(\mathcal{H}_S),\,\omega _\mu \in \mathcal{D}(%
\mathcal{H}_E)$ and numbers $c_\mu \in \mathbf{R}$ which satisfy $\sum_\mu
\left| c_\mu \right| <\infty $ and $\sum_\mu c_\mu =\mathrm{tr}\,W=1$. As a
consequence of assumption 2) the space $\mathcal{H}_E$ has infinite
dimension. If $\mathcal{H}_S$ is finite dimensional, the set (\ref{mod.18})
of statistical operators covers the whole space $\mathcal{D}(\mathcal{H}%
_{S+E})$. If also $\mathcal{H}_S$ is infinite dimensional, this set is dense
in $\mathcal{D}(\mathcal{H}_{S+E})$. With the arguments given above for
factorizing initial states the statement of Definition \ref{ss} can be
derived for all initial states (\ref{mod.18}), and the sectors $P_n\mathcal{H%
}_S$ are induced superselection sectors in the sense of this definition.
Moreover, assuming that the components of the statistical operator $W$
affiliated to the environment are sufficiently smooth functions in the
spectral representation of $V_E$, the sum $\sum_\mu \left| c_\mu \,\mathrm{tr%
}_E\left( \mathrm{e}^{-\imath (\lambda _m-\lambda _n)V_Et}\omega _\mu
\right) \right| $ satisfies a uniform estimate (\ref{mod.17}), and (\ref
{mod.1}) is still valid. Hence the time scale of the decoherence can be as
short as we want without restriction on $\rho (0)=\mathrm{tr}_EW=\sum_\mu
c_\mu \,\rho _\mu $.

If the potential $V_S$ has a (partially) continuous spectrum with spectral
family \\$\left\{ P(\Delta ),\,\Delta \subset \mathbf{R}\right\} $, an
estimate 
\begin{equation}
\left\| P(\Delta _1)\rho (t)P(\Delta _2)\right\| _2\leq C_\gamma (1+\delta
\left| t\right| )^{-\gamma }  \label{mod.1c}
\end{equation}
can be derived in the weaker Hilbert-Schmidt norm for arbitrary intervals $%
\Delta _1$ and $\Delta _2$ which have a non-vanishing distance, see Sect.
7.6 of \cite{GJKKSZ:1996}.

\subsection{The interaction with free fields: the role of infrared
divergence for induced superselection sectors\label{free}}

In this section we give up the restriction 3) on the Hamiltonian. Then the
estimate of the trace (\ref{mod.13}) needs more involved calculations. As
specific example we consider an environment given by a free Boson field.
Such models can be calculated explicitly, and they have often been used as
the starting point for Markov approximations.

As Hilbert space $\mathcal{H}_E$ we choose the Fock space based on the one
particle space $\mathcal{H}^{(1)}=\mathcal{L}^2(\mathbf{R}_{+})$ with inner
product $\left\langle f\mid g\right\rangle =\int_0^\infty \overline{f(k)}%
g(k)dk$. The one-particle Hamilton operator, denoted by $\widehat{%
\varepsilon }$, is the multiplication operator $\left( \widehat{\varepsilon }%
f\right) (k):=\varepsilon (k)f(k)$ with the energy function $\ \varepsilon
(k)=c\cdot k,\,c>0,\,k\in \mathbf{R}_{+}$, defined for all functions $f$
with $(1+\varepsilon (k))f(k)\in \mathcal{L}^2(\mathbf{R}_{+})$. The
creation/annihilation operators $a_k^{+}$ and $a_k$ are normalized to $%
\left[ a_k,a_{k^{\prime }}^{+}\right] =\delta (k-k^{\prime })$. The
Hamiltonian of the environment is then 
\begin{equation}
H_E=\int_0^\infty \varepsilon (k)a_k^{+}a_kdk.  \label{free.1}
\end{equation}
With $a^{+}(f)=\int_0^\infty f(k)a_k^{+}dk$ and $a(f)=\int_0^\infty
f(k)a_kdk $ we define field operators by $\Phi (f):=2^{-\frac 12}\left(
a^{+}(f)+a(f)\right) $ for real functions $f\in \mathcal{L}^2(\mathbf{R}%
_{+}) $. The interaction potential is chosen as $V_E=\Phi (f)$ with 
\begin{equation}
f\in \mathcal{L}^2(\mathbf{R}_{+})\mbox{  and }\,\widehat{\varepsilon }%
^{-1}f\in \mathcal{L}^2(\mathbf{R}_{+}),  \label{free.2a}
\end{equation}

An example for the total Hamiltonian is given by a single particle coupled
to the quantum field with velocity coupling 
\begin{equation}
\begin{array}{ll}
H_{S+E} & =\frac 12P^2\otimes I_E+P\otimes \Phi (f)+I_S\otimes H_E \\ 
& =\frac 12\left( P\otimes I_E+I_S\otimes \Phi (f)\right) ^2+I_S\otimes
\left( H_E-\frac 12\Phi ^2(f)\right)
\end{array}
\label{free.3}
\end{equation}
If the test function $f$ satisfies $\left\| \widehat{\varepsilon }^{-\frac
12}f\right\| <2^{-\frac 12}$, the Hamiltonian $H_E-\frac 12\Phi ^2(f)$ is
bounded from below, and consequently $H_{S+E}$ is bounded from below. Since
the particle is coupled to the free field with $V_S=P$, the reduced dynamics
yields continuous superselection sectors for the momentum $P$ of the
particle.

The operators (\ref{mod.8}) $\Gamma _m$ are substituted by $H_\lambda
:=H_E+\lambda \Phi (f),\;\lambda \in \mathbf{R}$, which are Hamiltonians of
the van Hove model \cite{Hove:1952}. The restrictions (\ref{free.2a}) are
necessary to guarantee that all operators $H_\lambda ,\,\lambda \in \mathbf{%
R,}$ are unitarily equivalent and defined on the same domain. To derive
induced superselection sectors we have to estimate the time dependence of
the traces $\chi _{\alpha \beta }(t):=\mathrm{tr}_EU_{\alpha \beta
}(t)\omega ,\,\alpha \neq \beta ,$ where the unitary operators $U_{\alpha
\beta }(t)$ are given by 
\begin{equation}
U_{\alpha \beta }(t):=\exp (iH_\alpha t)\exp (-iH_\beta t),  \label{free.4}
\end{equation}
see (\ref{mod.13}). In the Appendix we prove the following results for
states $\omega $ which are mixtures of coherent states.

\begin{enumerate}
\item[a)]  Under the restrictions (\ref{free.2a}) the traces $\chi _{\alpha
\beta }(t),\,\alpha \neq \beta ,$ do not vanish for $t\rightarrow \infty .$

\item[b)]  If $\Phi (f)$ has contributions at arbitrarily small energies, $%
\chi _{\alpha \beta }(t)$ can nevertheless strongly decrease for $\alpha
\neq \beta $ within a very long time interval $0\leq t\leq T$. Estimates
like (\ref{mod.1}) or (\ref{mod.1c}) are substituted by $\left\| P_m\rho
(t)P_n\right\| \leq f(t)$ or $\left\| P(\Delta )\rho (t)P(\Delta ^{\prime
})\right\| _2\leq f(t)$. But in contrast to (\ref{mod.1}) or (\ref{mod.1c})
the function $f(t)$ increases again for $t>T$.

\item[c)]  For fixed $\alpha \neq \beta $ a limit $\chi _{\alpha \beta
}(t)\rightarrow 0$ for $t\rightarrow \infty $ is possible if $\widehat{%
\varepsilon }^{-1}f\in \mathcal{L}^2(\mathbf{R}_{+})$ is violated, i.e. in
the case of infrared divergence.
\end{enumerate}

\noindent A large infrared contribution is therefore essential for the
emergence of induced superselection sectors. As in Sect. \ref{Araki} the
choice of the initial state $W$ of the total system can be extended to (\ref
{mod.18}) with $\rho _\mu \in \mathcal{D}(\mathcal{H}_S)$ and mixtures of
coherent states $\omega _\mu \in \mathcal{D}(\mathcal{H}_E)$. This class of
states is again dense in $\mathcal{D}(\mathcal{H}_{S+E})$, and, at least in
the infrared divergent case, we obtain induced superselection sectors in the
sense of Definition \ref{ss}.

\subsection{Models with scattering\label{scatt}}

For the models presented in Sects. \ref{Araki} and \ref{free} the projection
operators onto the effective superselection sectors $P_m\otimes I_S$ (or $%
P(\Delta )\otimes I_S$) commute with the total Hamiltonian. We now modify
the Hamiltonian (\ref{mod.2}) to 
\[
H=H_{S+E}+V=H_S\otimes I_E+I_S\otimes H_E+V_S\otimes V_E+V 
\]
where the operator $V$ is only restricted to be a \textit{scattering}
potential. This restriction means that the wave operator $\Omega
=\lim_{t\rightarrow \infty }e^{iHt}e^{-iH_{S+E}t}$ exists as strong limit.
To simplify the arguments we assume that there are no bound states such that
the convergence is guaranteed on $\mathcal{H}_{S+E}$ with $\Omega
^{+}=\Omega ^{-1}$. Then the time evolution $U(t)=\exp (-iHt)$ behaves
asymptotically as $U_0(t)\Omega ^{+}$ with $U_0(t)=\exp (-iH_{S+E}t).$ More
precisely, we have for all $W\in \mathcal{D(H}_{S+E}\mathcal{)}$%
\begin{equation}
\lim_{t\rightarrow \infty }\,\left\| U(t)WU^{+}(t)-U_0(t)\Omega ^{+}W\Omega
U_0^{+}(t)\right\| _1=0  \label{scatt.1}
\end{equation}
in trace norm. Following Sect. \ref{Araki} the reduced trace $\mathrm{tr}%
_EU_0(t)\Omega ^{+}W\Omega U_0^{+}(t)$ produces the superselection sectors $%
P_m\mathcal{H}_S$ which are determined by the spectrum (\ref{mod.5}) of $V_S$%
. The asymptotics (\ref{scatt.1}) then yields (in the sense of Definition 
\ref{ss}) the same superselection sectors for $\rho (t)=\mathrm{tr}%
_EU(t)WU^{+}(t)$. Moreover we can derive fast decoherence by additional
assumptions on the initial state and on the potential. For that purpose we
start with a factorizing initial state $W=\rho (0)\otimes \omega $ with
smooth $\omega .$ To apply the arguments of Sect. \ref{Araki} to the
dynamics $U_0(t)\Omega ^{+}W\Omega U_0^{+}(t)$ the statistical operator $%
\Omega ^{+}\left( \rho \otimes \omega \right) \Omega $ has to be a
sufficiently smooth operator on the tensor factor $\mathcal{H}_E$ for all $%
\rho \in \mathcal{D}(\mathcal{H}_S)$. That is guaranteed if we choose as
scattering potential a smooth potential in the sense of Kato \cite{Kato:1966}%
. Then both the limits, (\ref{scatt.1}) and\\$\lim_{t\rightarrow \infty
}\left\| P_m\left( \mathrm{tr}_EU_0(t)\Omega ^{+}W\Omega U_0^{+}(t)\right)
P_n\right\| _1=0,\,m\neq n,$ are reached in sufficiently short time. Hence $%
\rho (t)$ can decohere fast into the subspaces $P_m\mathcal{H}_S$ which are
determined by the spectrum (\ref{mod.5}) of $V_S$. But in contrast to (\ref
{mod.1}) one does not obtain a uniform bound with respect to the initial
state $\rho (0)$, since the limit (\ref{scatt.1}) is not uniform in $W\in 
\mathcal{D(H}_{S+E}\mathcal{)}$. \\Remark. The restriction that $V$ is a
scattering potential is essential. The dominating part $V_S\otimes V_E$ of
the interaction $V_S\otimes V_E+V$ still satisfies the assumption 1). In 
\cite{Kupsch:1996a} a spin model with an interaction which violates both the
constraints, the assumption 1) and the scattering condition, has been
investigated. That model can produce superselection sectors only in an
approximative sense, where the lower bounds on $\left\| P_m\rho
(t)P_n\right\| _1$ depend on the magnitude of the non-vanishing commutator.

\subsection{Concluding remarks}

The investigation of the models proves that the uniform emergence (\ref
{mod.1}) or (\ref{mod.1c}) of effective superselection sectors is consistent
with the mathematical rules of quantum mechanics. But this result depends on
rather restrictive assumptions on the Hamiltonian. For the more realistic
model of a quantum field presented in Sect. \ref{free} the suppression
persists only for a finite period of time. If the low frequency spectrum
dominates, this period of time can be sufficiently large for all practical
purposes. Only in the limit of infrared divergence the induced
superselection sectors persist for $t\rightarrow \infty $. If there is
additional scattering as considered in Sect. \ref{scatt} the superselection
sectors still exist. But the estimates are no longer uniform in the initial
state $\rho (0)$ of the system.

For all these models the induced superselection sectors are fully determined
by the Hamiltonian in the sense of Definition \ref{ss}. The initial state of
the total system, especially the smoothness properties of the components
related to the environment, determine the time scale in which these sectors
emerge.

\appendix

\section{The van Hove model\label{infra}}

As Hilbert space $\mathcal{H}_E$ we take the Fock space $\mathcal{F}(%
\mathcal{H}^{(1)})$ based on the one-particle space $\mathcal{H}^{(1)}=%
\mathcal{L}^2(\mathbf{R}_{+})$. For test functions $f,g\in \mathcal{S}(%
\mathbf{R}_{+})\subset \mathcal{L}^2(\mathbf{R}_{+})$ the creation and
annihilation operators $a^{+}(f)=\int_0^\infty f(k)a^{+}(k)dk$ and $%
a(g)=\int_0^\infty g(k)a(k)dk$ are normalized to $\left[
a(f),a^{+}(g)\right] =\left\langle f\mid g\right\rangle =\int_0^\infty 
\overline{f(k)}g(k)dk$. The test functions $f\in \mathcal{S}(\mathbf{R}_{+})$
are rapidly decreasing $C^\infty $-functions with a support restricted to $%
\mathbf{R}_{+}=\left[ 0,\infty \right) $. The one-particle Hamiltonian of
the free field is $\left( \widehat{\varepsilon }f\right) (k):=\varepsilon
(k)f(k)$ with the energy function $\ \varepsilon (k)=c\cdot k,\,c>0,$ for $%
k\geq 0$. Actually we can choose any positive monotonically increasing and
polynomially bounded energy function $\varepsilon (k)$, which has
excitations of arbitrarily small energy, $\varepsilon (k)/\left| k\right|
\rightarrow c>0$ if $k\rightarrow 0$. The Hamiltonian of the free field is
then (\ref{free.1}), and as canonical field and momentum operators we choose 
$\Phi (f):=\frac 1{\sqrt{2}}\left( a^{+}(f)+a(f)\right) $ and $\Pi
(f):=\frac i{\sqrt{2}}\left( a^{+}(f)-a(f)\right) .$ For real test functions
we define the Weyl operators 
\begin{equation}
T(f,g):=\exp \left( -i\Pi (f)-i\Phi (g)\right) =\exp \left( -i\Pi (f)\right)
\exp \left( -i\Phi (g)\right) \mathrm{e}^{-i\left( f\mid g\right) /2}
\label{f.4}
\end{equation}
These operators satisfy the Weyl relations 
\begin{equation}
T(f_1,g_1)T(f_2,g_2)=T(f_1+f_2,g_1+g_2)\mathrm{e}^{i\left( (f_1\mid
g_2)-(f_2\mid g_1)\right) /2}  \label{f.5}
\end{equation}
and their expectation value in the vacuum state $\Omega $ is 
\begin{equation}
\left\langle \Omega \mid T(f,g)\Omega \right\rangle =\exp \left( -\frac
14\left\| f\right\| ^2-\frac 14\left\| g\right\| ^2\right) .  \label{f.5a}
\end{equation}
With $U(t)=\exp (-iH_Et)$ the time evolution of the Weyl operators is 
\begin{equation}
U(-t)T(f,g)U(t)=T\left( \cos (\widehat{\varepsilon }t)\,f+\sin (\widehat{%
\varepsilon }t)\,g,\,\cos (\widehat{\varepsilon }t)\,g-\sin (\widehat{%
\varepsilon }t)\,f\right) .  \label{f.6}
\end{equation}
If the one-particle Hilbert space $\mathcal{L}^2(\mathbf{R}_{+})$ is
restricted to the one dimensional space $\mathbf{C,}$ all these formulas
become formulas of the one dimensional harmonic oscillator of frequency $%
\widehat{\varepsilon }=\ \varepsilon >0$.

The Weyl operator $T(f,0)$ is a translation operator 
\begin{eqnarray}
T(f,0)H_ET(-f,0) &=&\int_0^\infty \varepsilon (k)\left( a^{+}(k)+\frac 1{%
\sqrt{2}}f(k)\right) \left( a(k)+\frac 1{\sqrt{2}}f(k)\right) dk  \nonumber
\\
\ &=&H_E+\Phi (\widehat{\varepsilon }f)+\frac 12\int \varepsilon (k)\left|
f(k)\right| ^2dk.
\end{eqnarray}
Hence $T(\widehat{\varepsilon }^{-1}f,0)H_ET(-\widehat{\varepsilon }%
^{-1}f,0)-\frac 12\left\| \widehat{\varepsilon }^{-\frac 12}f\right\|
^2=H_E+\Phi (f)$ is the Hamiltonian of the van Hove model \cite{Hove:1952},
see also \cite{Cook:1961}, \cite{Berezin:1966} p. 166ff, and \cite{Emch:1972}%
. The operator $T(\widehat{\varepsilon }^{-1}f,0)$ is well defined if \\$%
\widehat{\varepsilon }^{-1}f\in \mathcal{L}^2(\mathbf{R}_{+}).$ The operator 
$\Phi (f),\,f\in \mathcal{L}^2(\mathbf{R}_{+})$, is $H_E$-bounded with
relative bound smaller than one (in the sense of the Kato-Rellich theorem,
see e.g. \cite{Reed/Simon:1975}) if $\widehat{\varepsilon }^{-\frac 12}f\in 
\mathcal{L}^2(\mathbf{R}_{+})$, and $\Phi ^2(f)$ is $H_E$-bounded with
relative bound smaller than one, if in addition $\left\| \widehat{%
\varepsilon }^{-\frac 12}f\right\| <2^{-1}$ holds. Hence the operators $H_E$
and $H_E+\lambda \Phi (f),\,\lambda \in \mathbf{R}$, are self-adjoint on the
same domain of the Fock space if $f\in \mathcal{L}^2(\mathbf{R}_{+})$ and $%
\widehat{\varepsilon }^{-\frac 12}f\in \mathcal{L}^2(\mathbf{R}_{+})$, and
moreover, the operator $H_E-\frac 12\Phi ^2(f)$ is bounded from below, if $%
\left\| \widehat{\varepsilon }^{-\frac 12}f\right\| <2^{-\frac 12}$.

The trace (\ref{mod.13}) is now calculated for the model of Sect. \ref{free}
with the one parameter family of Hamiltonians 
\begin{equation}
H_\lambda :=T(\lambda \widehat{\varepsilon }^{-1}f,0)H_ET(-\lambda \widehat{%
\varepsilon }^{-1}f,0)-\frac{\lambda ^2}2\left\| \widehat{\varepsilon }%
^{-\frac 12}f\right\| ^2=H_E+\lambda \Phi (f),\;\lambda \in \mathbf{R.}
\end{equation}
As stated above these operators are well defined if (\ref{free.2a}) holds.
In the following $\simeq $ indicates an identity up to a phase factor. The
unitary operators (\ref{free.3}) can be evaluated with the help of the Weyl
relations (\ref{f.5}) and the time evolution (\ref{f.6}) 
\[
\begin{array}{ll}
U_{\alpha \beta }(t) & \simeq T(\alpha \widehat{\varepsilon }%
^{-1}f,0)U(-t)T((\beta -\alpha )\widehat{\varepsilon }^{-1}f,0)U(t)T(-\beta 
\widehat{\varepsilon }^{-1}f,0) \\ 
& \simeq T\left( (\alpha -\beta )\widehat{\varepsilon }^{-1}(1-\cos \widehat{%
\varepsilon }t)\,f,\,(\alpha -\beta )(\widehat{\varepsilon }^{-1}\sin 
\widehat{\varepsilon }t)\,f\right) .
\end{array}
\]
We only investigate the trace $\mathrm{tr}_EU_{\alpha \beta }(t)\omega $ for
states $\omega $ which are mixtures of coherent states. Then the traces
decompose into sums of matrix elements of (\ref{free.3}) between coherent
states $T(f_n,g_n)\Omega $ with $f_n,\,g_n\in \mathcal{L}^2(\mathbf{R}%
_{+}),\,n=1,2,...$. These matrix elements can be calculated with the help of
(\ref{f.6}) and the Weyl relations (\ref{f.5}). Following (\ref{f.5a}) the
modulus of a matrix element is an exponential of the type 
\begin{equation}
\exp \left( -\frac 14\left\| a+(\alpha -\beta )(1-\cos \widehat{\varepsilon }%
t)\left( \widehat{\varepsilon }^{-1}f\right) \right\| ^2-\frac 14\left\|
b+(\alpha -\beta )\left( \sin \widehat{\varepsilon }t\right) \left( \widehat{%
\varepsilon }^{-1}f\right) \right\| ^2\right)  \label{f.13}
\end{equation}
where $a=f_m-f_n\in \mathcal{L}^2(\mathbf{R}_{+})$ and $b=g_m-g_n\in 
\mathcal{L}^2(\mathbf{R}_{+})$ are fixed. Under the restrictions (\ref
{free.2a}) the norms of $(1-\cos \widehat{\varepsilon }t)\left( \widehat{%
\varepsilon }^{-1}f\right) $ and $\left( \sin \widehat{\varepsilon }t\right)
\left( \widehat{\varepsilon }^{-1}f\right) $ are uniformly bounded, and (\ref
{f.13}) cannot vanish for $t\rightarrow \infty $.

But nevertheless, since $\varepsilon ^{-2}(1-\cos \varepsilon t)\rightarrow
\frac 12t^2$ and $\varepsilon ^{-1}\sin \varepsilon t\rightarrow t$ if $%
\varepsilon \rightarrow 0$, the norms $\left\| (1-\cos \widehat{\varepsilon }%
t)\left( \widehat{\varepsilon }^{-1}f\right) \right\| $ and $\left\| \left(
\sin \widehat{\varepsilon }t\right) \left( \widehat{\varepsilon }%
^{-1}f\right) \right\| $ may become as large as we want at intermediate
times, if $f(k)$ has large contributions at small values of $k$. The
estimate (\ref{f.13}) for the matrix elements is then negligible for a long
period of time (for arbitrary vectors $a$ and $b$ within some bounded
domain).

Only if we give up the second constraint in (\ref{free.2a}), we can find
test functions $f$ such that the norms of $(1-\cos \widehat{\varepsilon }%
t)\left( \widehat{\varepsilon }^{-1}f\right) $ and $\left( \sin \widehat{%
\varepsilon }t\right) \left( \widehat{\varepsilon }^{-1}f\right) $ increase
indefinitely for $t\rightarrow \infty $ and (\ref{f.13}) vanishes in this
limit.

That behaviour can be illustrated by the coupling to a free particle. As
already mentioned we can restrict the one-particle space $\mathcal{H}^{(1)}$
to the one dimensional space $\mathbf{C}$, and the free field becomes a
harmonic oscillator of frequency $\widehat{\varepsilon }=\varepsilon >0$. In
that case (\ref{f.13}) is a periodic function of $t\in \mathbf{R}$. In the
(singular) limit $\varepsilon \rightarrow 0$ we obtain functions $%
\varepsilon ^{-2}(1-\cos \varepsilon t)\rightarrow \frac 12t^2$ and $%
(\varepsilon ^{-1}\sin \varepsilon t)\rightarrow t$ which increase beyond
any bound for $t\rightarrow \infty .$ This limit case corresponds to the
Hamiltonian of a free particle 
\begin{equation}
H_E=\frac 12P^2\mbox{  with coupling }\,V_E=Q,  \label{f.15}
\end{equation}
and $\mathrm{tr}_EU_{\alpha \beta }(t)\omega $ can be calculated by standard
methods, see the article \cite{Pfeifer:1980} of Pfeifer, who has used this
model to discuss the measurement process of a spin. With (\ref{f.15}) the
Hamiltonian (\ref{mod.2}) of the total system is unbounded from below
(corresponding to infrared divergence in the field theoretic model) and we
have $\mathrm{tr}_EU_{\alpha \beta }(t)\omega \rightarrow 0$ if $%
t\rightarrow \infty $ for $\alpha \neq \beta $ and for all statistical
operators $\omega $ of the free particle.


\begin{thebibliography}{10}

\bibitem{Araki:1980}
H.~Araki.
\newblock {A remark on Machida-Namiki theory of measurement}.
\newblock {\em Prog. Theor. Phys.}, 64:719--730, 1980.

\bibitem{Berezin:1966}
F.~A. Berezin.
\newblock {\em The Method of Second Quantization}.
\newblock Academic Press, New York, 1966.

\bibitem{BLOT:1990}
N.~N. Bogolubov, A.~A. Logunov, A.~I. Oksak, and I.~T. Todorov.
\newblock {\em {General Principles of Quantum Field Theory}}.
\newblock Kluwer, Dortrecht, 1990.

\bibitem{Cook:1961}
J.~M. Cook.
\newblock {Asymptotic properties of a Boson field with given source}.
\newblock {\em J. Math. Phys.}, 2:33--45, 1961.

\bibitem{Emch:1972}
G.~G. Emch.
\newblock {\em {Algebraic Methods in Statistical Mechanics and Quantum Field
  Theory}}.
\newblock Wiley-Interscience, New York, 1972.

\bibitem{Emch:1972a}
G.~G. Emch.
\newblock On quantum measurement processes.
\newblock {\em Helv. Phys. Acta}, 45:1049--1056, 1972.

\bibitem{GJKKSZ:1996}
D.~Giulini, E.~Joos, C.~Kiefer, J.~Kupsch, I.~O. Stamatescu, and H.~D. Zeh.
\newblock {\em {Decoherence and the Appearance of a Classical World in Quantum
  Theory}}.
\newblock Springer, Berlin, 1996.

\bibitem{Hove:1952}
L.~{van} Hove.
\newblock Les difficult{\'e}s de divergences pour un mod{\`e}le particulier de
  champ quantifi{\'e}.
\newblock {\em Physica}, 18:145--159, 1952.

\bibitem{Joos/Zeh:1985}
E.~Joos and H.~D. Zeh.
\newblock The emergence of classical properties through interaction with the
  environment.
\newblock {\em Z. Phys.}, B59:223--243, 1985.

\bibitem{Kato:1966}
T.~Kato.
\newblock Wave operators and similarity for some non-selfadjoint operators.
\newblock {\em Math. Annalen}, 162:258--279, 1966.

\bibitem{Kupsch:1996a}
J.~Kupsch.
\newblock The structure of the quantum mechanical state space and induced
  superselection rules.
\newblock {Lecture at the Workshop on Foundations of Quantum Theory, T.I.F.R.
  Bombay}, 1996.
\newblock quant-ph/9612033.

\bibitem{Mirman:1979}
R.~Mirman.
\newblock Nonexistence of superselection rules: Definition of term {\it frame
  of reference}.
\newblock {\em Found. Phys.}, 9:283--299, 1979.

\bibitem{Peres:1995}
A.~Peres.
\newblock {\em {Quantum Theory: Concepts and Methods}}.
\newblock Kluwer, Dordrecht, 1995.

\bibitem{Pfeifer:1980}
P.~Pfeifer.
\newblock A simple model for irreversible dynamics from unitary time evolution.
\newblock {\em Helv. Phys. Acta}, 53:410--415, 1980.

\bibitem{Piron:1969}
C.~Piron.
\newblock Les r{\'e}gles de supers{\'e}lection continues.
\newblock {\em Helv. Phys. Acta}, 42:330--338, 1969.

\bibitem{Reed/Simon:1975}
M.~Reed and B.~Simon.
\newblock {\em {Methods of Modern Mathematical Physics II, Fourier Analysis:
  Self-Adjointness}}.
\newblock Academic Press, New York, 1975.

\bibitem{Wightman:1995}
A.~S. Wightman.
\newblock Superselection rules; old and new.
\newblock {\em Nuovo Cimento}, 110B:751--769, 1995.

\bibitem{Zeh:1970}
H.~D. Zeh.
\newblock On the interpretation of measurement in quantum theory.
\newblock {\em Found. Phys.}, 1:69--76, 1970.

\bibitem{Zeh:1971}
H.~D. Zeh.
\newblock On irreversibility of time and observation in quantum theory.
\newblock In B.~D'Espagnat, editor, {\em {Foundations of Quantum Mechanics}},
  pages 263--273, New York, 1971. Academic Press.

\bibitem{Zurek:1982}
W.~H. Zurek.
\newblock Environment induced superselection rules.
\newblock {\em Phys. Rev.}, D26:1862--1880, 1982.

\end{thebibliography}
\end{document}